\documentclass{acm_proc_article-sp}

\newcommand{\ans}[3]{\mathrm{ans}(#1,#2,#3)} 

 \renewcommand{\P}{\mathcal{P}}
 \newcommand{\R}{\mathcal{R}}
\renewcommand{\S}{\mathcal{S}}

\newcommand{\la}{\leftarrow}

\newcounter{cefalo}
\newcounter{cefalocont}

\newtheorem{theorem}{Theorem} 

\newtheorem{definitionAux}[theorem]{Definition}

\newtheorem{claimAux}{Claim}

\newtheorem{exampleAux}{Example} 

\newtheorem{examplesAux}[theorem]{Examples}

\newtheorem{constructionAux}[theorem]{Construction}

\def\qed{\hfill{\qedboxempty}      
  \ifdim\lastskip<\medskipamount \removelastskip\penalty55\medskip\fi}

\def\qedboxempty{\vbox{\hrule\hbox{\vrule\kern3pt
                 \vbox{\kern3pt\kern3pt}\kern3pt\vrule}\hrule}}

\def\qedfull{\hfill{\qedboxfull}   
  \ifdim\lastskip<\medskipamount \removelastskip\penalty55\medskip\fi}

\def\qedboxfull{\vrule height 4pt width 4pt depth 0pt}

\newcommand{\markfull}{\qedfull}

\newcommand{{\incolumn}}[1]{\begin{tabular}[c]{c} #1 \end{tabular}}
\newcommand{{\incolumnmath}}[1]{\begin{array}[c]{c} #1 \end{array}}

\usepackage{enumerate}
\usepackage[defblank]{paralist}

\begin{document}

\title{Determining Relevant Relations for Datalog Queries under Access Limitations is Undecidable}
\subtitle{[Extended Abstract]}
%
%
%
%
%

\numberofauthors{1} 
%
\author{
%
%
\alignauthor
Davide Martinenghi\\
       \affaddr{Politecnico di Milano}\\
       \affaddr{Piazza Leonardo 32}\\
       \affaddr{20132 Milano, Italy}\\
       \email{davide.martinenghi@polimi.it}
}

\maketitle
\begin{abstract}
Access limitations are restrictions in the way in which the tuples of a relation can be accessed. Under access limitations, query answering becomes more complex than in the traditional case, with no guarantee that the answer tuples that can be extracted (aka \emph{maximal answer}) are all those that would be found without access limitations (aka \emph{complete answer}).
The field of query answering under access limitations has been broadly investigated in the past.
Attention has been devoted to the problem of determining relations that are \emph{relevant} for a query, i.e., those (possibly off-query) relations that might need to be accessed in order to find all tuples in the maximal answer.
In this short paper, we show that relevance is undecidable for Datalog queries.
\end{abstract}

%
%
%
%
%
%
%
%
\section{Relevance}

The problem of querying data sources that have limited capabilities and can thus only be accessed by providing values for certain fields according to given patterns has raised a great deal of interest in the past few years~\cite{Hale01,Levy99,Li03,MiLF00,DuLe97,FLMS99,LiCh00,LiCh01,M:FQAS2004,LiCh01b,CM:LOPSTR2003,Deutsch:2007lr,DM:FlexDBIST2007,DBLP:journals/jcss/MillsteinHF03,DM:FlexDBIST2006,YaKC06,CM:AAI2000,NaLu04}.

An access pattern is a constraint indicating which attributes of a relation schema are used as input and which ones are used as output.

In this respect, access patterns may suitably characterize several relevant contexts, such as Web forms, legacy data, Web services, and the so-called Deep Web~\cite{CM:APWEB2010,CM:EDBT2010,M:CRYPT2011}. Query processing under access patterns requires specialized techniques.
Among these, static optimization, including query containment, has been studied for several forms of conjunctive queries and unions thereof~\cite{CM:ER2008,CM:ICDE2008,CCM:EROW2007,CMC:SEBD2007,LiCh01b,DBLP:conf/pods/BenediktGS11}. More general cases are covered in the context of dynamic optimization~\cite{CCM:JUCS2009}, where results are available for schemata with functional dependencies and simple full-width inclusion dependencies. The latter kind of dependencies, albeit simple, can be used to state equivalence, and thus captures the notion of relations with multiple access patterns.

In the context of access limitations, it is important to distinguish between the maximal answer and the complete answer to a query.
The \emph{maximal answer} is the largest set of query answers that can be computed from the relations in the schema over which the query is posed, while complying with the access limitations.
The \emph{complete answer} to a query is the answer that could be computed if we could retrieve all the tuples from the relations in the query as if with no access limitations.
In the following, we indicate with $\ans{q}{\R}{D}$ the set of tuples in the maximal answer to a query $q$ over a schema $\R$ with access limitations on a database $D$.
The maximal answer can be computed by means of a Datalog program, as described in~\cite{DBLP:journals/jlp/DuschkaGL00}.

A relation $r$ in a relation schema $\R$ is \emph{relevant} for a query $q$ if there are two database instances $D$ and $D'$ over $\R$ 
for which $\ans{q}{\R}{D}\neq\ans{q}{\R}{D'}$
and such that $r^D\neq r^{D'}$ and $s^D=s^{D'}$ for every relation $s\neq r$ in $\R$.

We show that relevance is undecidable in the case of Datalog queries, since, if we were able to decide relevance of a relation, we could also decide containment between Datalog queries, which is known to be undecidable.

\begin{theorem}\label{the:relevant-datalog-undecidable}
	Testing relevance for Datalog queries is undecidable.
\end{theorem}
\begin{proof}
Let $\Pi$ be a Datalog program over a schema $\R$, without access limitations, defining two arbitrary predicates $p$ and $q$.
Let $e$ be an extensional predicate not occurring in $\Pi$ and $i$ a new intensional predicate defined by the rules 
$i(\vec X)\la e, p(\vec X)$, and $i(\vec X)\la q(\vec X)$.
If we could establish whether $e$ is relevant to answer the Datalog query
$Ans(\vec X) \la i(\vec X)$
then we could also decide containment between $p$ and $q$, which is absurd (and \emph{a fortiori} absurd for a schema that can have access limitations).
More precisely, $q$ contains $p$ iff $e$ is not relevant, i.e.:
\begin{enumerate}
\item[(1)] If $q$ contains $p$, then $e$ is not relevant.
\item[(2)] If $q$ does not contain $p$, then $e$ is relevant.
\end{enumerate}

To see that (1) holds, it suffices to observe that $p$ cannot contribute any tuple to $i$ that is not already contributed by $q$, and thus $e$ need not be accessed.

To see that (2) holds, consider a database $D$ in which $p(\vec c)$ holds but $q(\vec c)$ does not hold, for some constants $\vec c$. Such a database must exist, as $p$ is not contained in $q$.
Since $p$ and $q$ are independent of $e$, their containment is also independent of $e$, so $e$ may either hold or not hold in $D$. If $e$ holds in $D$, then $\vec c$ is in the answer; if $e$ does not hold in $D$, then $\vec c$ is not in the answer. Therefore $e$ is relevant, since it changes the maximal answer to the query.
\end{proof}

\bibliographystyle{abbrv}
\bibliography{short-string,POZZ2057,bibsections-eng,cv}  
\end{document}